\documentclass[aps,twocolumn,showpacs]{revtex4-1}
\usepackage{graphicx}
\usepackage{amsmath,amsfonts}
\usepackage[T1]{fontenc}

\begin{document}

\title{Dimensional crossover in a Fermi gas and a cross-dimensional Tomonaga-Luttinger model}

\author{Guillaume Lang}
\affiliation{Universit\'e Grenoble Alpes, LPMMC, F-38000 Grenoble, France}
\affiliation{CNRS, LPMMC, F-38000 Grenoble, France}
\author{Frank Hekking}
\affiliation{Universit\'e Grenoble Alpes, LPMMC, F-38000 Grenoble, France}
\affiliation{CNRS, LPMMC, F-38000 Grenoble, France}
\author{Anna Minguzzi}
\affiliation{Universit\'e Grenoble Alpes, LPMMC, F-38000 Grenoble, France}
\affiliation{CNRS, LPMMC, F-38000 Grenoble, France}

\begin{abstract}
We describe the dimensional crossover in a noninteracting Fermi gas in an anisotropic trap, obtained by populating various transverse modes of the trap. We study the dynamical structure factor and drag force. Starting from a dimension $d$, the $(\!d\!+\!1\!)$-dimensional case is obtained to a good approximation with relatively few modes. We show that the dynamical structure factor of a gas in a $d$-dimensional harmonic trap simulates an effective $2d$-dimensional box trap. We focus then on the experimentally relevant situation when only a portion of the gas in harmonic confinement is probed and give a condition to obtain the behavior of a $d$-dimensional gas in a box. Finally, we propose a generalized Tomonaga-Luttinger model for the multimode configuration and compare the dynamical structure factor in the 2D limit with the exact result, finding that it is accurate in the backscattering region and at low energy.
\end{abstract}

\maketitle

\section{Introduction and motivation}
In the last decades, low-dimensional systems of ultracold atoms have attracted attention with many achievements, such as the observation of the superfluid to Mott insulator transition \cite{Bloch}, solitons \cite{Fleischer}, the Berezinskii-Kosterlitz-Thouless transition \cite{Hadzi} and the realization of a Tonks-Girardeau gas \cite{Paredes}. In particular, the possibility to trap and confine ultracold atoms to various one-dimensional (1D) \cite{VanDruten, Roati, Billy, Esslinger, Chen, Naegerl} and two-dimensional (2D) geometries \cite{Dalibard} raises new questions regarding the dynamical behavior of such systems. Yet, few of them are really one- or two-dimensional. For a wide class of transverse confinements, they are better modeled as quasi-one- (Q1D) or quasi-two- (Q2D) dimensional, meaning that transverse modes can be populated \cite{Widera, Gorlitz, Chung}. Indeed, multimode systems are ubiquitous in condensed matter, in particular in electronic systems where multichannel quantum wires \cite{Sandler}, 2-leg ladders \cite{Crepin}, carbon nanotubes \cite{Gogolin} and biased bilayer graphene \cite{Killi} are a few examples.

Dimensional crossovers in Q1D and Q2D systems, e.g. from 1D to 2D or even 3D, can occur in at least two ways. The first one is the occupation of various energy modes in a given tight confining trap. The transition to higher dimension is realized by gradually opening the trap, thereby increasing the number of populated modes in energy space. The second scheme is the realization of a higher-dimensional structure by an ensemble of low-dimensional ones in real space, e.g. in an optical lattice \cite{Sommer, Joung, Vogler}.

Especially suitable quantities to probe dimensional crossovers in ultracold gases are the dynamical structure factor and the drag force since they are strongly dependent on the dimensionality of the system. The dynamical structure factor is of peculiar interest to characterize the dynamical response of the fluid to a moving potential barrier or an impurity, and is measurable by Bragg scattering experiments \cite{Stenger, Ozeri, Calabrese, David, Meinert, Brunello}. The drag force measures the heating rate during the process \cite{Dalibard, Ketterle, Singh, Pitaevskii, Brand}. In this paper, we provide calculations of these quantities for noninteracting fermions in experimental situations realizing a dimensional crossover in energy space. Some analytical expressions for the dynamical structure factor are already known in 1D and 3D \cite{Pines,Fetter}. We provide a more general expression depending explicitly on the dimension and show how the crossover occurs by considering an increasing number of occupied transverse energy modes. The effect of an external longitudinal trapping confinement is known to considerably change the dynamical structure factor \cite{Vignolo}. Within the local density approximation, we show that the harmonic trap enhances the effective dimension of the system, thus allowing to simulate the physics in a box trap up to six dimensions. Reciprocally, we show how to prevent this enhancement experimentally and simulate the dynamical structure factor in a 1D box-trap using an harmonically confined gas.

While non-interacting fermions are amenable to exact calculations, they are also an ideal testbed to develop new approximation schemes. In 1D, the low-energy excitations of the gas can be described using the Tomonaga-Luttinger model (TLM) \cite{Tomonaga, Mattis, Haldane, Cazalilla, Giamarchi}. We show how the well-known breakdown of the TLM in dimensions higher than one \cite{Voit} is reflected in the behavior of the dynamical structure factor and the drag force. We propose a multimode Tomonaga-Luttinger model (M-TLM) that correctly captures the behavior of higher dimensional systems near the backscattering region.

The paper is organized as follows: in Sec.~\ref{Syst} we use a multimode approach to show dimensional crossovers from 1D to $d$D in a box. In Sec.~\ref{trap} we show that, within the local density approximation, the $d$-dimensional dynamical structure factor in a harmonic trap has the same shape as in a $2d$-dimensional box. In Sec.~\ref{Luttinger}, we extend the Tomonaga-Luttinger model to multimode systems and dimensions higher than one. In Sec.~\ref{Outlook} we summarize the main results and give a few outlooks.

\section{Energy space dimensional crossover in a box trap}
\label{Syst}
\subsection{The system}

We consider $N$ ultracold non-interacting spinless fermions of mass $m$ in an anisotropic uniform box confinement at zero temperature. We assume that the length $L_x$ of the box is much longer than its width $L_y$ and height $L_z$. This situation can be approached experimentally in an optical box trap \cite{Hadzibabic}. If at least one of the transverse sizes is small enough, such that the level spacing is larger than all characteristic energy scales of the problem (e.g. temperature, chemical potential), then the gas is confined to 2D or even to 1D, since the occupation of higher transverse modes is suppressed. In the following, we study the behavior of the system as transverse sizes are gradually increased and the transverse modes occupied. This yields a dimensional crossover from 1D to 2D and eventually 3D.

\subsection{Dimensional crossover for the dynamical structure factor}
First, we are interested in the effect of the dimension on the dynamical structure factor. The latter contains all the information about the structure and collective excitations of the gas. In arbitrary dimension $d$ in a box-trap, it reads
\begin{equation}
\label{defS}
S_{d}(\vec{q},\omega)\!=\!V_d\!\!\int_{-\infty}^{+\infty}\!\!\!\!\!\!dt\!\int\!\! d^d\!r\ \!e^{i(\omega t-\vec{q}\cdot \vec{r})} \!\langle \delta n_d(\vec{r},t)\delta n_d(\vec{0},0)\rangle,
\end{equation}
where $V_d$ is the volume of the system, $\hbar \vec{q}$ and $\hbar\omega$ are the transferred momentum and energy in the Bragg spectroscopy process, $\delta n_d(r,t)\equiv n_d(\vec{r},t)-N/V_d$ are the fluctuations of the $d$-dimensional density operator at time $t$ and $\langle \dots \rangle$ stands for the equilibrium quantum statistical average.

Before discussing the dimensional crossover, we determine the dynamical structure factor of a $d$-dimensional gas in the thermodynamic limit, for $d=1,2,3$. We specialize to $\vec{q}=q\vec{e}_x$, where $\vec{e}_x$ is the unit vector along the $x$-axis. These results can be written in a compact form as a general power law which depends explicitly on $d$ and reads
\begin{eqnarray}
\label{Sd}
 &&S_{d}(q \vec{e}_x,\omega)=V_ds_{d}\left(\frac{m}{2\pi\hbar q}\right)^d\times \nonumber\\
 &&\left[\Theta(\omega_{+}\!-\!\omega)\Theta(\omega\!-\!\omega_{-})(\omega_{+}\!-\!\omega)^{\frac{d\!-\!1}{2}}(\omega\!-\!\alpha_d\omega_{-})^{\frac{d\!-\!1}{2}}\right.\nonumber\\
 &&+ \Theta(2k_{F}\!-\!q)\Theta(\omega_{-}\!-\!\omega)\nonumber\\
 &&\left.\left\{\![(\omega_{-\!}\!+\!\omega)(\omega_{+}\!-\!\omega)]^{\frac{d\!-\!1}{2}}\!\!\!-\![(\omega_{+}\!+\!\omega)(\omega_{-\!}\!-\!\omega)]^{\frac{d\!-\!1}{2}}\!\right\}\!\right]\!\!.
\end{eqnarray}
Here $\Theta$ is the Heaviside distribution, $k_{F}=\left[\frac{N}{V_d}\frac{(2\pi)^d}{\Omega_d}\right]^{1/d}$ is the modulus of the $d$-dimensional Fermi wavevector, where $\Omega_d\!=\!\pi^{\frac{d}{2}}/\Gamma(\frac{d+2}{2})$ is the volume of the unit $d$-dimensional ball with $\Gamma$ the Euler Gamma function, $\omega_{\pm}\equiv\left|\frac{\hbar q^2}{2m}\!\pm\!\frac{\hbar k_{F}q}{m}\right|$ are the boundaries of the energy-momentum sector where particle-hole excitations can occur, $\alpha_d\equiv$ sign$(q_x\!-\!2k_{F\!})$, and $s_d=2\pi^{\frac{d+1}{2}}\!/ \Gamma(\frac{d+1}{2})$ is the surface of the unit $d$-sphere.

We now consider the dimensional crossover of the dynamical structure factor from $S_{d}$ to $S_{d+1}$, obtained by populating higher transverse modes of the atomic waveguide. We illustrate this procedure by focusing on the dimensional crossover from dimension one to two in a 2D box with periodic boundary conditions.
We write the two-dimensional fermionic field operator as $\psi(x,y)=\sum_{k_x}\sum_{k_y}\frac{e^{ik_xx}}{\sqrt{L_x}}\frac{e^{ik_yy}}{\sqrt{L_y}}a_{k_xk_y}$, where $k_{x,y}\!=\!\frac{2\pi}{L_{x,y}}j_{x,y}$ with $j_{x,y}$ an integer, and $a_{k_xk_y}\!\equiv\!a_{\vec{k}}$ is the fermionic annihilation operator, such that $\{a_{\vec{k}},a^{\dagger}_{\vec{k}'}\}\!=\!\delta_{\vec{k},\vec{k}'}$, and $\langle a_{\vec{k}}^{\dagger}a_{\vec{k}'}\rangle=\delta_{\vec{k},\vec{k'}}n_F(\epsilon_k)$, where $n_F(\epsilon_k)\equiv \frac{1}{e^{\beta(\epsilon_k-\mu)}+1}$ is the Fermi-Dirac distribution, with $\beta$ the inverse temperature, $\mu$ the chemical potential, and $\epsilon_k\!=\!\frac{\hbar^2k^2}{2m}\equiv \hbar \omega_k$ the free-particle dispersion relation. Then, applying Wick's theorem we find that
\begin{eqnarray}
\label{Wick}
 \langle\delta n(\vec{r},t)n(\vec{0},0)\rangle&&=\!\frac{1}{L_x^2}\frac{1}{L_y^2}\sum_{\vec{k},\vec{k}'}\!e^{-i[(\vec{k}\!-\!\vec{k}')\cdot \vec{r}\!-\!(\omega_{k_x}\!+\!\omega_{k_y}\!-\!\omega_{k_x'}\!-\!\omega_{k_y'}\!)t]}\nonumber\\
 &&n_F(\epsilon_{k_x}\!+\!\epsilon_{k_y})[1-n_F(\epsilon_{k_x'}\!+\!\epsilon_{k_y'})].
\end{eqnarray}
Substituting Eq.(\ref{Wick}) into Eq.(\ref{defS}), the dynamical structure factor reads
\begin{eqnarray}
 &&S_{Q1}(q\vec{e}_x,\omega)\!=\!\frac{L_x}{L_y}\int_{-\infty}^{+\infty}\!\!\!dt\!\int_{-L_x/2}^{L_x/2}\!\!\!\!dx\int_{-L_y/2}^{L_y/2}\!dy e^{i(\omega t-q_x x)}\frac{1}{(2\pi)^2}\nonumber\\
 &&\!\!\int_{-\infty}^{+\infty}\!\!\!\!\!dk_x\!\!\!\int_{-\infty}^{+\infty}\!\!\!\!\!dk_x'\!\!\sum_{k_y,k_y'}\!\!\!e^{-i[(k_x\!-\!k_x')x+(k_y\!-\!k_y')y-\!(\omega_{k_x}\!+\!\omega_{k_y}\!-\!\omega_{k_x'}\!-\!\omega_{k_y'}\!)t]}\nonumber\\
 &&n_F(\epsilon_{k_x}\!+\!\epsilon_{k_y})[1-n_F(\epsilon_{k_x'}\!+\!\epsilon_{k_y'})].
 \end{eqnarray}
 A few additional algebraic manipulations and specialization to $T=0$ yield
 \begin{eqnarray}
 \label{Equa5}
 &&S_{Q1}(q\vec{e}_x,\omega)=\sum_{k_y}\!2\pi L_x\!\!\int_{-\infty}^{+\infty}\!dk_x \nonumber\\
 &&\Theta[\epsilon_F\!-\!(\epsilon_{k_x}\!+\!\epsilon_{k_y})]\Theta[\epsilon_{k_x\!+\!q_x}\!+\!\epsilon_{k_y}\!-\!\epsilon_F]\delta[\omega\!-\!(\omega_{k_x+q_x}\!-\!\omega_{k_x})]\nonumber\\
 &&=\sum_{j_y=-M}^MS_1(q\vec{e}_x,\omega;\tilde{k}_{F}[j_y/M]),
\end{eqnarray}
where $\epsilon_F\!=\!\mu\!=\!\frac{\hbar^2k_F^2}{2m}$ is the Fermi energy, $S_1(q\vec{e}_x,\omega;\tilde{k}_F[j_y/M])$ is the 1D dynamical structure factor where the chemical potential has been replaced by $\epsilon_F\!-\!\epsilon_{k_y}$, or equivalently, where the wavevector $k_{F,1}$ has been replaced by $\tilde{k}_{F}[j_y/M]\equiv k_{F}\sqrt{1\!-\!\frac{j_y^2}{\tilde{M}^2}}$, which defines the number of transverse modes $2M+1$ by $M=I[\tilde{M}]$, where $\tilde{M}\equiv \frac{k_FL_y}{2\pi}$ and $I$ is the integer part function. Finally, in the large $M$ limit, the Riemann sum in Eq.~(\ref{Equa5}) becomes an integral, and one obtains
\begin{eqnarray}
\label{crossdim}
S_2(q\vec{e}_x,\omega)=\frac{k_{F}L_y}{\pi}\int_0^1dx S_1(q\vec{e_x},\omega;k_F\sqrt{1-x^2}),
\end{eqnarray}
providing the dimensional crossover from 1D to 2D.

More generally, one can start from a system of any dimension $d$ and find, after relaxation of the transverse confinement,
\begin{eqnarray}
\label{induc}
&&S_{Qd}(q\vec{e}_x,\omega)=\sum_{j=-M}^MS_d(q\vec{e}_x,\omega;\tilde{k}_{F}[j/M])_{\stackrel{\longrightarrow}{_{M\to +\infty}}}\nonumber\\
&&\frac{k_{F}L_{d+1}}{\pi}\!\!\int_0^1\!\!dx S_d(q\vec{e_x},\omega;\tilde{k}_{F}[x])=S_{d+1}(q\vec{e}_x,\omega).
\end{eqnarray}
If used repeatedly, Eq.~(\ref{induc}) allows to compute the dynamical structure factor up to any dimension if it is known in lower dimension. In particular, it allowed us to prove Eq.~(\ref{Sd}) by induction. A detailed illustration of the crossover from 1D to 2D can be found in Appendix \ref{d=2}, generalizations to any integer dimension rely on the same techniques.

The derivation of Eq.(\ref{induc}) also shows that
\begin{eqnarray}
\label{dirS}
&&S_d(q\vec{e}_x,\omega)=V_d \int \frac{d^dk}{(2\pi)^{d-1}}\nonumber\\
&&\Theta\!\!\left(\!\!\epsilon_F\!-\!\!\!\sum_{i=1}^d\!\!\epsilon_{k_{x_i}}\!\!\!\right)\!\!\Theta\!\!\left(\!\sum_{i=1}^{d}\epsilon_{k_{x_i}\!+\!q\delta_{i,1}}\!\!\!\!-\!\epsilon_F\!\!\right)\!\delta[\omega\!-\!\!(\omega_{k_{x_1}\!+\!q}\!\!-\!\omega_{k_{x_1}}\!)]
\end{eqnarray}
in agreement with the expression found by using Lindhard's formula \cite{Ashcroft} for the density-density response function combined with the fluctuation-dissipation theorem. We have cross-checked our results by using Eq.~(\ref{induc}) and Eq.~(\ref{dirS}) independently to compute the dynamical structure factor in 2D and 3D.

\begin{figure}
\includegraphics[width=8.5cm, keepaspectratio, angle=0]{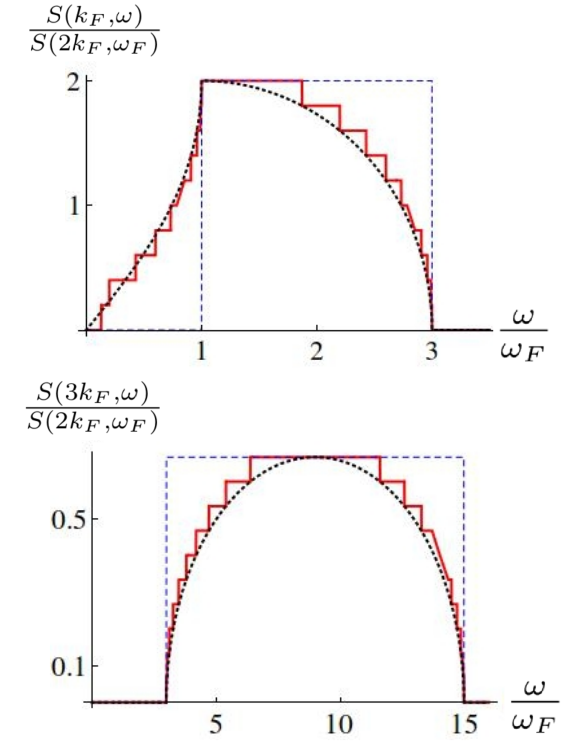}%10modesS(omega)q1.jpeg}
\caption{(Color online) Dynamical structure factor $S(q,\omega)$ in units of $S(q\!=\!2k_F,\omega\!=\!\omega_F)$ for dimensionless wavevectors $q/k_F\!=\!1$ (upper panel) and $q/k_F\!=\!3$ (lower panel), with $k_F$ the Fermi momentum, as a function of frequency $\omega$ in units of the Fermi frequency $\omega_F\equiv \frac{\hbar k_F^2}{2m}$, in 1D (blue, dashed) and Q1D for $2M\!+\!1\!=\!21$ modes (red, solid) compared to 2D (black, dotted). Few modes are needed for the Q1D system to display a similar behavior as the 2D one.}
\label{Sq13}
\end{figure}

Now, we illustrate numerically the dimensional crossover from 1D to 2D using Eq.~(\ref{induc}). Figure~\ref{Sq13} shows the dynamical structure factor as a function of the frequency $\omega$ for two choices of wavevector $q$. In each panel are represented a 1D gas, a Q1D gas for $M=10$ and the 2D result for a comparison. Sections are made at fixed $q$ rather than $\omega$ because they are obtained in experiments \cite{David, Meinert}. We notice that only a few modes are needed to recover within a very good approximation the higher-dimensional physics, since in this example, the staircase shape taken by the dynamical structure factor of the Q1D gas mimics already quite well the 2D one. We have checked that this is also the case in the 1D to 3D and in the 2D to 3D crossovers.

\subsection{Dimensional crossover for the drag force}
\label{crossF}
The dimensional crossover is a powerful approach which allows to derive the drag force in arbitrary dimension $d$.
In detail, from the dynamical structure factor one can extract information about the dynamical response of the fluid to a small perturbation. If a weak potential barrier or impurity is stirred along the fluid, putting it slightly out of equilibrium, then in linear response theory the average energy dissipation per unit time is linked to the dynamical structure factor by the relation \cite{Pitaevskii}
\begin{eqnarray}
&&\langle\overline{\dot{E}}\rangle\!=\!-\frac{1}{2\pi\hbar V_d}\!\int_0^{+\infty}\!\!\!\!d\omega\!\! \int \!\!\frac{d^dq}{(2\pi)^d} S_{d}(\vec{q},\omega)|U_{d}(\vec{q},\omega)|^2\omega,
\end{eqnarray}
where $U_{d}(\vec{q},\omega)$ is the Fourier transform of the potential barrier $U_d(\vec{r},t)$ defining the perturbation part of the Hamiltonian $H_{pert}\equiv \int d^dr U_d(\vec{r},t)n_d(\vec{r})$.

In analogy with classical hydrodynamics, the concept of drag force is introduced to quantify the viscosity, and defined as $\langle\overline{\dot{E}}\rangle\!\equiv\!-\!\vec{F}\cdot \vec{v}$, where $\langle\overline{\dot{E}}\rangle$ is the average heating rate per unit time, $\vec{v}$ is the velocity of the potential barrier, which we assume constant in the following. In the context of ultracold atoms, measuring the heating rate is a way to probe superfluidity \cite{Dalibard, Ketterle}. Indeed, a non-viscous flow leads to a vanishing drag force at low enough velocity, thus $F\!=\!0$ is a necessary condition for superfluidity, usually called drag force criterion.

With a delta-potential barrier $U_d(\vec{r},t)\!\equiv\!U_d \delta(x\!-\!vt)$ in the direction $x$, covering the whole waveguide in the transverse directions, the drag force at zero temperature reads
\begin{eqnarray}
\label{explF}
 F_{d}(v)=\frac{U_d^2}{2\pi\hbar V_d}\int_0^{+\infty}\!\!\!dqS_{d}(q\vec{e_x},qv)q.
\end{eqnarray}

In 1D we find, in agreement with \cite{Pitaevskii,Cherny}, that
\begin{equation}
F_{1}(v)\!=\!\frac{2U_1^2mn_1}{\hbar^2}\!\left[\Theta(v\!-\!v_{F,1})\!+\!\frac{v}{v_{F,1}}\Theta(v_{F,1}\!-\!v)\right]\!,
\end{equation}
where $v_{F,d}\!\equiv\!\frac{\hbar k_{F,d}}{m}$ is the Fermi velocity. We also compute the drag force in 2D and 3D. The results found when $v\leq v_{F,d}$ are
\begin{eqnarray}
\label{F2}
\!\!\!\!\!\!\!\!F_{2}(v)\!=\!\!\frac{2U_2^2mn_{2}}{\hbar^2}\frac{2}{\pi}\!\!\left[\!\frac{v}{v_{F,2}}\!\sqrt{1\!-\!\left(\!\frac{v}{v_{F,2}}\!\right)^2}\!+\!\arcsin\!\left(\!\frac{v}{v_{F,2}}\!\right)\!\!\right]
\end{eqnarray}
and
\begin{eqnarray}
F_{3}(v)=\frac{2U_3^2mn_{3}}{\hbar^2}\frac{3}{2}\frac{v}{v_{F,3}}\left[1-\frac{1}{3}\left(\frac{v}{v_{F,3}}\right)^2\right].
\end{eqnarray}
If $v\!>\!v_{F,d}$, for the potential barrier considered, the drag force saturates at the universal value $F_{d}(v_{F,d})=2U_d^2mn_{d}/\hbar^2$. We recovered those results by applying the cross-dimensional approach from dimension $d$ to dimension $(d\!+\!1)$, which validates this technique once more, as illustrated in Appendix~\ref{d=2} in the case $d\!=\!1$.
%$U_2^2\leftrightarrow \frac{k_{F,2}}{2\pi}g_i^2$ and $U_3^2\leftrightarrow \frac{k_{F,3}^2}{(2\pi)^2}g_i^2$, which is compatible with the fact that dimensionally $[U_d\delta^d]=[g_i\delta^1]$.

From these expressions it is not easy to guess a general formula for any integer dimension $d$. Carrying out the calculation from Eqs.~(\ref{Sd}) and (\ref{explF}), we found that it actually reads
\begin{eqnarray}
\label{Fdd}
&&F_d(u_d\leq 1)=C_d(1-u_d^2)^{\frac{d-1}{2}} \nonumber\\
&&\left[\!(1\!+\!u_d){_2}F_1\!\!\left(\!1,\!\frac{1\!-\!d}{2};\!\frac{d\!+\!3}{2};\!-\frac{1\!+\!u_d}{1\!-\!u_d}\right)\!-\!(\!u_d\!\rightarrow\!-u_d)\!\right]\!\!,
\end{eqnarray}
where $C_d\!\equiv\!\frac{2U_d^2mn_d}{\hbar^2}\frac{2}{\sqrt{\pi}(d+1)}\frac{\Gamma(\frac{d+2}{2})}{\Gamma(\frac{d+1}{2})}$ is a numerical coefficient, ${_2}F_1(a,b;c;x)\!\equiv\!\sum_{n=0}^{+\infty}\frac{(a)_n(b)_n}{(c)_n}\frac{x^n}{n!}$ is the hypergeometric function with $(a)_n\equiv \prod_{i=0}^{n-1}(a+i)$ the Pochhammer symbol, and we introduced the notation $u_d\equiv \frac{v}{v_{F,d}}$. In integer dimensions the hypergeometric function reduces to simple functions, for a technical discussion and expressions which do not involve special functions, we refer to Appendix~\ref{Fd}.

\begin{figure}
\includegraphics[width=8cm, keepaspectratio, angle=0]{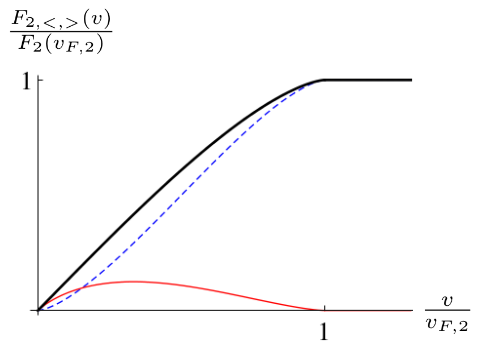}%F.jpeg}
\caption{(Color online) Two-dimensional drag force $F_2$ (black,thick), $F_<$ (dashed, blue) and $F_>$ (red) in units of $F_2(v_{F})$, with $v_F$ the Fermi velocity, as functions of the velocity $v$ in units of $v_F$.}
\label{Breakforce}
\end{figure}

To gain some insight in the structure of these results we focus on the 2D situation and split the drag force Eq.~(\ref{F2}) onto two contributions, one due to the integration of the dynamical structure factor below $\omega_-$, which we call $F_<$, while the other, called $F_>$, is due to the contributions above $\omega_-$.
Then, the drag force for $v<v_F$ is $F_2\equiv F_{2,>}+F_{2,<}$ with
\begin{eqnarray}
\label{Fsup}
\frac{F_{2,>}(u)}{F_2(u=1)}=\frac{2}{\pi}\left[\arcsin\left(\sqrt{\frac{1+u}{2}}\right)+f_-(u)\right]
\end{eqnarray}
and
\begin{eqnarray}
\label{Finf}
\!\!\!\!\!\!\frac{F_{2,<}(u)}{F_2(u\!=\!1)}\!=\!\frac{2}{\pi}\!\!\left[u\sqrt{1\!-\!u^2}\!-\!\arcsin\!\left(\!\!\sqrt{\frac{1\!-\!u}{2}}\!\right)\!-\!\!f_-(u)\!\right]\!,
\end{eqnarray}
where $f_-(u)\!\equiv\!(2u\!-\!1)\sqrt{u(1\!-\!u)}\!+\!\arctan\left(\sqrt{\frac{u}{1\!-\!u}}\right)\!-\!\arctan\left(\sqrt{\frac{1\!+\!u}{1\!-\!u}}\right)$ is the boundary term at $\omega_-$. As can be seen in Fig.\ref{Breakforce}, $F_<$ and $F_>$ are of the same order of magnitude even for $v\ll v_F$, showing that the continuum of particle-hole excitations below $\omega_-$ and above $\omega_-$ contribute equally to the drag force. This is very different from the one-dimensional case, which sustains particle-hole excitations only for energies above $\omega_-$. As we will discuss in more details in Sec.~\ref{Luttinger} below, these results raise the question of the generalization of the 1D Tomonaga-Luttinger model (TLM) to higher dimensions. We will show that the dimensional crossover discussed above provides a valuable route to generalize the TLM to dimensions higher than one.

According to the drag force criterion, the noninteracting Fermi gas is not superfluid, as expected since superfluidity is a collective phenomenon, requiring the presence of interactions. One of the first theoretical insights of an interacting case was provided in \cite{Pitaevskii}, with a potential barrier $U_d(\vec{r},t)\!\equiv\!g_i\delta(\vec{r}-\vec{v}t)$ in a weakly-interacting Bose gas. A mean-field approach was used, neglecting solitons and vortices. The various results in dimension $d\in\{1,2,3\}$ can be recast into the general form
\begin{equation}
 F_{d}(v)=\frac{s_{d-1}}{(2\pi)^{d-1}}\frac{m^dg_i^2n_{d}}{\hbar^{d+1}}\left(\frac{v^2-c_d^2}{v}\right)^{d-1}\Theta(v-c_d),
\end{equation}
where $c_d$ is a critical velocity, coinciding with the sound velocity in the mean-field approach. It is interesting that dimensionality manifests in a simple power law, even in the presence of interactions. It is even simpler than for free fermions, due to the richer structure of the dynamical structure factor for the latter.

To conclude this Section, we want to emphasize that the dimensional crossover approach provides a variety of angles to attack a given problem. We illustrate this point in Appendix \ref{d=2} in dimension $d=2$, where the drag force is found in three different ways.

\section{Dimensional crossovers in a harmonic trap}
\label{trap}
After having discussed the dimensional crossover in energy space in a box trap, we focus here on dimensional crossovers in the experimentally relevant situation of a harmonically trapped gas.

\subsection{Equivalence between a $d$-dimensional ideal gas in a harmonic trap and a $2d$-dimensional ideal gas in a box}

We consider a 1D Fermi gas longitudinally confined by a harmonic trap described by the potential $V(x)=\frac{1}{2}m\omega_{0}^2x^2$, where $\omega_{0}$ is the frequency of the trap. Assuming a slow spatial variation along $x$ allows us to use the local density approximation (LDA) to describe the density profile of the gas. To this end, we describe the system by a position-dependent chemical potential according to the relation
\begin{eqnarray}
 \mu-\frac{1}{2}m\omega_0^2x^2=\mu_{hom}[n(x)],
\end{eqnarray}
where $\mu_{hom}[n(x)]=\frac{\hbar^2\pi^2n^2}{2m}$ is the equation of state for a 1D homogeneous Fermi gas. Combined with the normalization condition $\int_{-R_{TF}}^{R_{TF}}dx\ n(x)=N$, this yields
\begin{eqnarray}
\label{dom}
 n(x)=\frac{2N}{\pi R_{TF}}\sqrt{1-\frac{x^2}{R_{TF}^2}}\Theta(R_{TF}-|x|).
\end{eqnarray}
where $R_{TF}\!\equiv\!\sqrt{\frac{2\mu}{m\omega_{0}^2}}$ is the Thomas-Fermi radius.

Within the same approximation of a slowly varying spatial confinement we calculate the dynamical structure factor $S_{1,HO}^{LDA}(q,\omega)$ of the harmonically trapped gas. In detail, for wavevectors $q$ larger than the inverse scale of the spatial confinement $1/R_{TF}$ we take the spatial average
\begin{eqnarray}
S_{1,HO}^{LDA}(q,\omega)=\!\!\frac{1}{2R_{TF}}\!\!\int_{-R_{TF}}^{R_{TF}}\!\!dx S_{1,hom}[q,\omega;n(x)],
\end{eqnarray}
where $S_{1,hom}[q,\omega;n]$ is the dynamical structure factor of a 1D homogeneous gas of density $n$. This local density approximation assumes that portions of the size of the confinement length scale $a_{H0}=\sqrt{\frac{\hbar}{m\omega_0}}$ can be considered as homogeneous and that their responses are independent from each other \cite{Golovach}. The validity of this approximation has been verified in \cite{Vignolo} by comparing it with exact results.

After the change of variable $x/R_{TF}\to x$, we find
\begin{eqnarray}
\label{SLDA}
S_{1,HO}^{LDA}(q,\omega)=\int_0^1dxS_1(q,\omega;n_1\sqrt{1-x^2})
\end{eqnarray}
with $n_1\!=\!\frac{2N}{\pi R_{TF}}$. It has the same form as Eq.(\ref{crossdim}), thus establishing the equivalence, in terms of the dynamical structure factor, of a 1D harmonic trapped gas and a 2D gas in a box.

%\begin{figure}
%\includegraphics[width=1.7cm, keepaspectratio, angle=0]{syst1D.png}
%\includegraphics[width=2.5cm, keepaspectratio, angle=0]{systq1D.png}
%\includegraphics[width=3cm, keepaspectratio, angle=0]{syst2D.png}
%\caption{(Color online) Illustration of the dimensional crossover from a single uniform 1D gas (left) to a 2D uniform gas (right), expected as the 1D system is duplicated conserving the total number of atoms (center) in the limit of large number of 1D systems. Colors are an indication of the expected density in a box, light tones standing for the lower values.}
%\label{cross}
%\end{figure}

\begin{figure}
\includegraphics[width=8.5cm, keepaspectratio, angle=0]{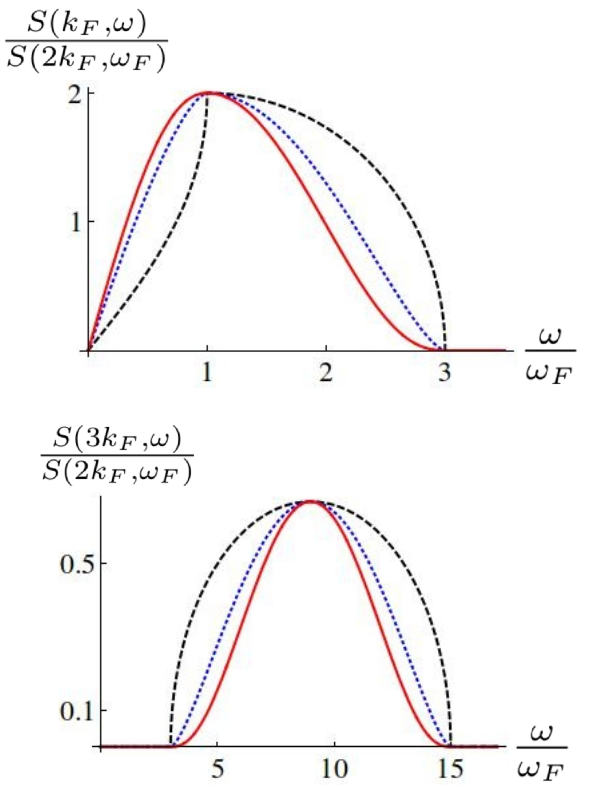}%S36Dq1.jpeg}
\caption{(Color online) Simulated dynamical structure factor $S(q,\omega)$ in units of $S(q\!=\!2k_F,\omega\!=\!\omega_F)$ for dimensionless wavevectors $q/k_F\!=\!1$ (upper panel) and $q/k_F\!=\!3$ (lower panel), with $k_F$ the Fermi momentum, as a function of frequency $\omega$ in units of the Fermi frequency $\omega_F\equiv \frac{\hbar k_F^2}{2m}$, in 2D (black, dashed), 4D (blue, dotted) and 6D (red, solid). }
\label{S36D}
\end{figure}

So far we have assumed a strictly 1D geometry. More generally, a similar procedure allows to obtain one of the main results of this Section: in reduced units, the dynamical structure factor of a harmonically trapped ideal gas in $d$D is the same as in a box trap in 2$d$D.
We illustrate in Fig.\ref{S36D} the dynamical structure factor of an ideal gas in a box in dimensions $d\in\{2,4,6\}$ as can be simulated by a harmonically confined ideal gas in dimension $d=1,2,3$ respectively.

The correspondence between a 2$d$D box trap and a $d$D harmonic trap can be inferred directly from the Hamiltonian of the system: for a box trap there are $d$ quadratic contributions originating from the kinetic energy whereas for the harmonic confinement there are $2d$ quadratic terms coming from both kinetic and potential energy. Since, in a semiclassical treatment each term contributes in a similar manner, harmonic confinement leads to a doubling of the effective dimensionality of the system in the non-interacting case, which is expected not only for the dynamical structure factor, but also for other quantities such as the density of states, or the condensed fraction of a trapped Bose gas below the critical temperature for instance.

\subsection{Interpolation from 2D box behavior to 1D box behavior in a harmonic trap}

We have pointed out how a harmonic trap increases the effective dimension of the gas for the dynamical structure factor. Conversely, we analyse now how the dynamical structure factor of a harmonically confined gas looks like if only the central part of the cloud is probed over a length $r<R_{TF}$. Assuming that $r$ is larger than the characteristic length of the variation of the external confinement, and again using the local density approximation, Eq.~(\ref{SLDA}) reduces to
\begin{eqnarray}
 S_{1,HO}^{LDA}(q,\omega;r)\simeq \int_{0}^{r/R_{TF}}\!\!\!dxS_{1}(q,\omega;n_1\sqrt{1-x^2}).
\end{eqnarray}
An explicit expression is obtained by evaluating the integral
\begin{eqnarray}
\!\!\!\!\!\!I\!\!\equiv\!\!\!\!\int_0^{r/R_{TF}}\!\!\!\!\!\!\!\!\!\!\!\!\!dx \Theta(q^2\!+\!2q\sqrt{\!1\!-\!x^2}\!-\!\omega)\Theta(\omega\!-\!|q^2\!-\!2q\sqrt{\!1\!-\!x^2}|)
\end{eqnarray}
where $\omega$ and $q$ are expressed in reduced units such that $k_F=1$ and $\omega_F=1$. The final expression reads
\begin{eqnarray}
\label{I}
 &&I=\Theta(\omega_+\!-\!\omega)\Theta(\omega\!-\!\omega_-)\min\!\left(\!\frac{r}{R_{TF}},\sqrt{1-\left(\frac{\omega-q^2}{2q}\right)^2}\right)\nonumber\\
 &&+\Theta(2\!-\!q)\Theta(\omega_-\!-\!\omega)\min\!\left(\!\frac{r}{R_{TF}},\sqrt{1-\left(\frac{\omega-q^2}{2q}\right)^2}\right)\nonumber\\
 &&-\Theta(2\!-\!q)\Theta(\omega_-\!-\!\omega)\min\!\left(\!\frac{r}{R_{TF}},\sqrt{1-\left(\frac{\omega+q^2}{2q}\right)^2}\right)\!\!.
\end{eqnarray}

\begin{figure}
\includegraphics[width=8.5cm, keepaspectratio, angle=0]{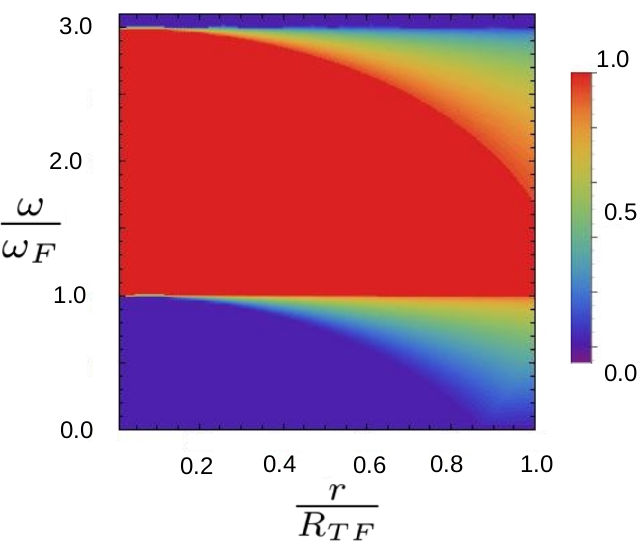}%reffect.jpeg}
\caption{(Color online) Reduced dynamical structure factor $S(q=k_F,\omega;r)/r$ in units of $S_1(q\!=\!k_F,\omega)$ in the plane $(r,\omega)$, where $r$ is the probed length of the gas in units of $R_{TF}$ and $\omega$ the frequency in units of $\omega_F$. If $r\ll R_{TF}$ one recovers the 1D box result, while $r\to R_{TF}$ yields the $2D$ box result. Excitations below the lower branch appear progressively as the dimensionless ratio $r/R_{TF}$ is increased.}
\label{r}
\end{figure}

This expression displays the crossover between the behavior of a 1D gas in a box and the one of a 2D gas in a box. In order to obtain the 1D behavior, $\frac{r}{R_{TF}}$ must be the minimal argument in Eq.~(\ref{I}) above, while the 2D behavior is obtained when $\frac{r}{R_{TF}}$ is the largest argument. In essence, to get close to the 1D behavior one should take the smallest $r$ compatible with the condition $r\gtrsim \frac{1}{q}$ assumed in order to use the LDA, and with the condition $1-\frac{r}{R_{TF}}\ll 1$ in order to detect enough signal. Figure \ref{r} shows the dynamical structure factor $S_{1,HO}^{LDA}(q,\omega;r)$ as a function of frequency, at varying the size $r$ of the probe region, for a fixed choice $q=k_F$. Quite remarkably, we find that, for our choice of $q$, the 1D behavior is probed to a large accuracy up to a size $r\lesssim 0.3R_{TF}$.

\section{Low-energy approach for fermions in a box trap}
\label{Luttinger}

\subsection{The Tomonaga-Luttinger model in $d=1$}

In 1D, gapless systems with linear dispersion relation lie within the universality class of the Tomonaga-Luttinger model (TLM), associated with the Hamiltonian
\begin{eqnarray}
 H^{TL}=\frac{\hbar v_s}{2\pi}\int dx\left[K(\partial_x\phi)^2+\frac{1}{K}(\partial_x\theta)^2\right],
\end{eqnarray}
where TL stands for Tomonaga-Luttinger, involving canonically conjugate fields such that $[\partial_x\theta(x),\phi(x')]=i\pi\delta(x-x')$, the dimensionless Luttinger parameter $K$ and the sound velocity $v_{s}$.
At low energy, the dispersion relation of the free Fermi gas in a box can be linearized around the Fermi points $\pm k_F$. With this simplification, after some algebraic manipulations \cite{Haldane}, one finds that the low-energy effective Hamiltonian is described by the TLM with parameters $K=1$ and $v_s=v_F$. This provides us with an alternative formalism to study the dynamical structure factor and the drag force. Although in the present work we will focus on the noninteracting limit, where it will be compared to the exact solution, this formalism can describe interacting systems as well and is worth studying in this perspective.

As far as the dynamical structure factor is concerned, the effective theory yields the linearization of its definition domain at the origin and around the umklapp point $(q\!=\!2k_F,\omega\!=\!0)$ which corresponds to backscattering processes, intrinsically limiting its domain of validity to low energies (for more details, see e.g.~\cite{Lang} and references therein).
Since we specialize to noninteracting fermions, the dynamical structure factor in the backscattering region reads
\begin{eqnarray}
\label{STLM}
\!\!\!S_{1}^{TL}(q,\omega)=L_x B_1(K=1)\Theta(\omega-|q-2k_F|v_F),
\end{eqnarray}
where $B_1$ is a model-dependent coefficient which depends on $K$. Comparison with the exact result at the umklapp point yields $B_1(K\!=\!1)\!=\!\frac{m}{2\hbar k_{F}}$. Then, Eq.~(\ref{STLM}) reproduces the exact dynamical structure factor given by Eq.~(\ref{Sd}) with less than $10\%$ error, due to the linearization, provided that $\omega \lesssim 0.3\omega_F$ \cite{Lang}.

\subsection{Generalized Tomonaga-Luttinger model for $d\!>\!1$}

\begin{figure}
\includegraphics[width=7.5cm, keepaspectratio, angle=0]{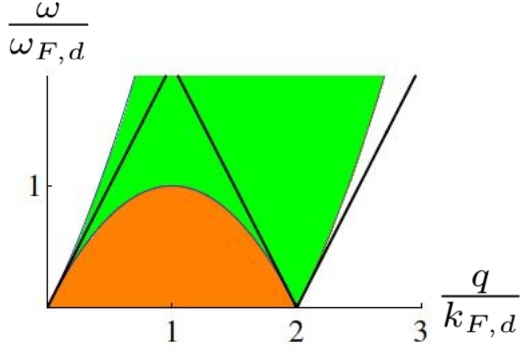}%modes.jpeg}
\caption{(Color online) Definition domain of the dynamical structure factor of a Fermi gas in the plane $(\omega,q)$ in units of $(\omega_{F,d},q_{F,d})$. Shaded areas represent the domain where single particle-hole excitations can occur. The light green one is found in any integer dimension $d\in \{1,2,3\}$, while the dark orange one is specific to $d>1$. Black straight lines correspond to the linearization of the domain in the Tomonaga-Luttinger formalism in 1D.}
\label{SdD}
\end{figure}

One can see in Fig.~\ref{SdD} and in Eq.~(\ref{Sd}) that in 2D and 3D, since excitations are possible at energies below $\omega_{-}$ up to $\omega=0$ for any $q<2k_F$, no linearization of the dynamical structure factor is possible and the standard Tomonaga-Luttinger liquid theory breaks down. Actually, these excitations progressively appear as soon as $d>1$ \cite{Metzner}, yet fractal geometries which would allow to probe noninteger dimensions are beyond the scope of this work.

\begin{figure}
\includegraphics[width=8cm, keepaspectratio, angle=0]{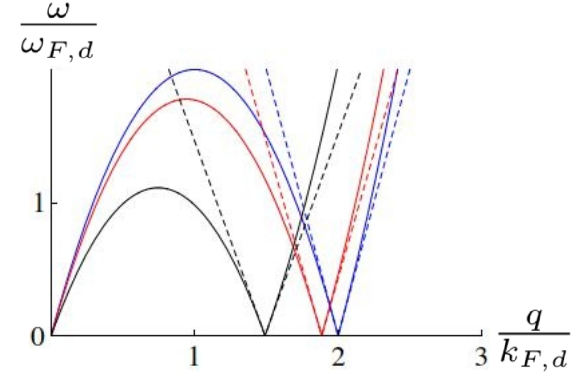}%multilin.jpeg}
\caption{(Color online) Lower boundary of the definition domain for the dynamic structure factor of a q1D gas with three modes, in the plane $(q,\omega)$  in units of $(k_F,\omega_F)$, as found in the Tomonaga-Luttinger formalism (dashed) compared to the exact solution (solid).}
\label{multilin}
\end{figure}

Many attempts were made to generalize the Tomonaga-Luttinger model to higher dimensions \cite{Wen, Bartosch} as an alternative to Fermi liquids to describe interacting systems. In this work, we use the crossover approach as in Sec.~\ref{Syst} to construct a Tomonaga-Luttinger model in higher dimensions, defining a multimode Tomonaga-Luttinger model (M-TLM). Indeed, the emergence of contributions at energies lower than $\omega_-$ in the dynamical structure factor if $d>1$ can be interpreted as contributions of higher modes of a 1D gas. All those modes, taken separately, display a linear structure in their dynamical structure factor at low energy, as illustrated in Fig.\ref{multilin}. Thus, applying Eq.~(\ref{crossdim}) to the Tomonaga-Luttinger model, in Q1D the dynamical structure factor reads
\begin{eqnarray}
\label{multiLutt}
S^{TL}_{Q1}(q,\omega)=\sum_{j=-M}^MS_{1}^{TL}(q,\omega;\tilde{k}_F[j/M]).
\end{eqnarray}

The question is, up to what point the small errors for each mode in the framework of the effective theory amplify or cancel when adding more modes if compared to the exact result, especially in the limit $M\!\to\!+\!\infty$ which corresponds to the crossover to 2D. To address this question, we carry out the procedure explicitly on the example of the 1D to 2D crossover and compare the prediction of our theory to the exact solution. We combine Eq.~(\ref{multiLutt}) to Eq.~(\ref{STLM}) and find
\begin{eqnarray}
S^{TL}_{Q1}(q,\omega)&&=L_x\frac{m}{4\pi\hbar}\frac{1}{\tilde{M}}\sum_{j=-M}^M\frac{1}{\sqrt{1-\frac{j^2}{\tilde{M}^2}}}\times \nonumber\\
 &&\Theta\!\!\left(\!\omega\!-\!\left|q\!-\!2k_F\sqrt{1\!-\!\frac{j^2}{\tilde{M}^2}}\right|\!v_F\sqrt{1\!-\!\frac{j^2}{\tilde{M}^2}}\!\right)\nonumber\\
&&\to_{M\to +\infty}L_x\frac{m}{2\pi\hbar}\int_0^1 dx\frac{1}{\sqrt{1-x^2}}\times \nonumber\\
 &&\Theta\!\!\left(\!\omega\!-\!\left|q\!-\!2k_F\sqrt{1\!-\!x^2}\right|\!v_F\sqrt{1\!-\!x^2}\!\right).
\end{eqnarray}
Evaluating the integral yields
\begin{eqnarray}
\!\!\!\!\!\!\!\!\!S^{TL}_2(q,\omega)\!=\!\frac{mL_x}{2\pi\hbar}\!\left[\Theta(q\!-\!2k_F)S_>\!\!+\!\Theta(2k_F\!-\!q)S_<\right]\!(q,\omega)
\end{eqnarray}
with
\begin{eqnarray}
\label{Sup}
&&S_>(q,\omega)=\Theta\left(\frac{\hbar q^2}{8m}-\omega\right)\Theta(\tilde{q}v_F-\omega)\nonumber\\
&&\arcsin\left(\frac{q}{4k_F}\left[1-\sqrt{1-\frac{8m\omega}{\hbar q^2}}\right]\right)\nonumber\\
&&+\Theta(4k_F-q)\Theta(\omega-\tilde{q}v_F)\Theta\left(\frac{\hbar q^2}{8m}-\omega\right)\nonumber\\
&&\!\arcsin\!\left[\!\frac{q}{4k_F}\!\!\left(\!1\!-\!\!\sqrt{1\!-\!\frac{8m\omega}{\hbar q^2}}\!\right)\!\right]\!\!+\!\arccos\!\left[\frac{q}{4k_F}\!\!\left(1\!+\!\!\sqrt{1\!-\!\frac{8m\omega}{\hbar q^2}}\right)\!\right]\nonumber\\
&&+\Theta\left(\omega-\frac{\hbar q^2}{8m}\right)\Theta(\omega-\tilde{q}v_F)\frac{\pi}{2}
\end{eqnarray}
and 
\begin{eqnarray}
\label{Sinf}
&&S_<(q,\omega)=\Theta(|\tilde{q}|v_F-\omega)\arcsin\left(\frac{q}{4k_F}\left[1+\sqrt{1+\frac{8m\omega}{\hbar q^2}}\right]\right)\nonumber\\
&&+\Theta\left(\frac{\hbar q^2}{8m}-\omega\right)\nonumber\\
&&\!\arcsin\!\left[\!\frac{q}{4k_F}\!\!\left(\!1\!-\!\!\sqrt{1\!-\!\frac{8m\omega}{\hbar q^2}}\!\right)\!\right]\!\!-\!\arcsin\!\left[\frac{q}{4k_F}\!\!\left(1\!+\!\!\sqrt{1\!-\!\frac{8m\omega}{\hbar q^2}}\right)\!\right]\nonumber\\
&&+\Theta(\omega-|\tilde{q}|v_F)\frac{\pi}{2},
\end{eqnarray}
where $\tilde{q}\equiv q-2k_F$.

\begin{figure}
\includegraphics[width=8.5cm, keepaspectratio, angle=0]{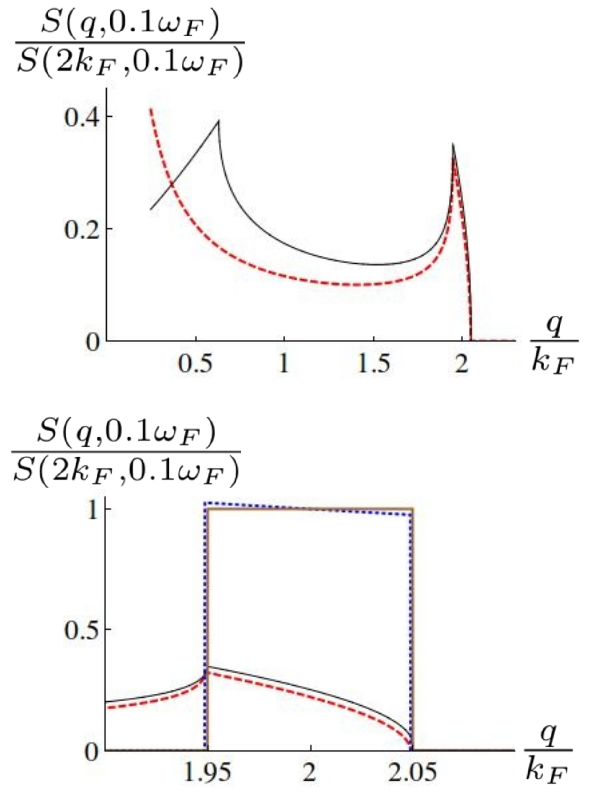}%Vs4.jpeg}
\caption{(Color online) Section of the dynamical structure factor $S(q,\omega\!=\!0.1\omega_F)$ in units of $S(q\!=\!2k_F,\omega\!=\!0.1\omega_F)$ as a function of $q$ in units of $k_F$, at fixed dimensionless frequency $\omega/\omega_F=0.1$. The exact result in 2D (red, dashed) is compared to the M-TLM prediction (solid, black) in the upper panel. The lower panel shows a zoom into the backscattering region near $q=2k_F$. It compares the 2D exact (red, dashed) and the M-TLM model (solid, black) to the exact (brown, thick) and TLM (dotted, blue) results in 1D.}
\label{Vs4}
\end{figure}

We illustrate Eqs.~(\ref{Sup}) and (\ref{Sinf}) in Fig.\ref{Vs4}, where we compare sections of the dynamical structure factor, as a function of $q$ at $\omega=0.1\omega_F$. Around the umklapp point $(q\!=\!2k_F,\omega\!=\!0)$ there is an area where the approximate model is in a rather good quantitative agreement with the exact result in 2D. The differences between the two models at low $q$ are due to the fact that for a given point, the TLM slightly overestimates the value of the dynamical structure factor for larger $q$ and underestimates it at lower $q$, as can be seen in the 1D case. Combined with the fact that the curvature of the dispersion relation has been neglected, and that the density of modes is lower at low $q$, this explains both the anomalous cusp and the falling down of the M-TLM at low $q$. Note however that the M-TLM result is by far closer to the 2D exact result than the 1D one, showing that there is a true multimode effect.

We have checked that the M-TLM predictions for a noninteracting gas are in quantitative agreement with the exact 2D result for $\omega\ll \omega_F$ and $|q-2k_F|\ll 2k_F$. Similar conditions have to be met in 1D in order to ensure the validity of the Tomonaga-Luttinger model \cite{Lang}, therefore our heuristic construction is quite satisfactory from this point of view.

\section{Summary and outlook}
\label{Outlook}

We have computed the dynamical structure factor and drag force of a noninteracting Fermi gas in an anisotropic box confinement in dimension $d=1,2,3$.
We have developed a multimode treatment in energy space and have pointed out the crossover from any integer dimension to a higher one. We have found that only a few modes are needed to recover higher-dimensional physics to a good approximation.

Using the local density approximation and a Thomas-Fermi profile for the density, we have shown that in a harmonic trap, each added degree of confinement is equivalent to increasing the effective dimension by one unit compared to a box trap for an ideal gas. This allows to simulate the dynamical structure factor of an ideal gas in a box up to six effective dimensions using a harmonic confinement. Reciprocally, probing a region around the center of the trap can reduce the effective dimension, allowing to extract experimentally the physics of the 1D ideal gas from a higher-dimensional one, and we have identified the conditions.

As far as low-energy excitations are concerned, the dynamical structure factor has allowed us to illustrate the breakdown of the standard Tomonaga-Luttinger model in dimensions higher than one. Yet, we have shown that a multimode approach provides a suitable generalization of the model to two and possibly higher dimensions around the backscattering region.

Although our results have been obtained for a non-interacting Fermi gas, we conjecture that the M-TLM could be able to give (at least qualitative) hints about the behavior of multimode or multicomponent interacting bosons or fermions, even in the presence of some types of couplings (e.g. density-density) in an array of 1D wires. Some analytical results are already available for two \cite{Iucci} or three \cite{Orignac} coupled Tomonaga-Luttinger liquids, but most challenging is the limit of an infinite number of components \cite{Guinea} as compared with higher-dimensional interacting systems. This may give new insights on the conditions to observe a transition from a 1D Luttinger liquid to a 2D Fermi liquid, studied in \cite{Metzner}, and more generally, on the huge differences between the low-dimensional and the 3D systems, for instance in the appearance of vortices or solitons.

Our M-TLM model could be further improved by taking into account the effects of the curvature (see e.g. \cite{Imambekov} for a review) and thermal excitations.

\acknowledgments

We thank David Clement for fruitful discussions.

We acknowledge financial support from ANR project Mathostaq (ANR-13-JS01-0005-01), ANR project SuperRing (ANR-15-CE30-0012-02), ERC Handy-Q grant N.258608 and from Institut Universitaire de France.

%%%%%%%%%%%%%%%%%%%%%%%%%%%%%%%%%%%%%%%%%%%%%%%%%%%%%%%%%%%%%%%%%%%%%%%%%%%%%%%%%%%%%%%%%%%%%%%%%%%%%%%%%%%%%%%%%%%%%%%
\appendix
%%%%%%%%%%%%%%%%%%%%%%%%%%%%%%%%%%%%%%%%%%%%%%%%%%%%%%%%%%%%%%%%%%%%%%%%%%%%%%%%%%%%%%%%%%%%%%%%%%%%%%%%%%%%%%%%%%%%%%%

\section{Comparison of three different approaches to find the drag force in dimension $d=2$}
\label{d=2}
In this Appendix we illustrate three approaches to compute the drag force for a given dimension, provided by the dimensional crossovers. We focus on the case of the drag force in $d=2$.

We first perform a direct calculation of both the dynamical structure factor and drag force in 2D, then we use the crossover approach to derive the dynamical structure factor, derive it in 2D and deduce from it the drag force in 2D. Finally, we calculate both the dynamical structure factor and drag force in 1D, and use the crossover approach to obtain the drag force in 2D.

\subsection{Approach (I): direct calculation of $S_2$ and $F_2$}
According to Eq.(\ref{dirS}), one has to calculate
\begin{eqnarray}
\label{A1}
&&S_2(q,\omega\!)\!=\!L_xL_y\frac{1}{2\pi}\!\!\int \!\!\mathrm{d}^2k\nonumber\\
&&\Theta(k_F\!-\!|k|)\Theta(|k\!+\!q|\!-\!k_F)\delta[\omega\!-\!(\omega_{k+q}\!-\!\omega_k)].
\end{eqnarray}
The main difficulty of the calculation consists in finding the integration boundaries. From a physical point of view, they are defined by an interplay between the energy conservation in the scattering process, described by the delta distribution, and the impenetrability of the two Fermi spheres (discs in $d\!=\!2$), described by the Heaviside distributions.

The task of finding the maximal integration domain allowed by the Heaviside distributions can be translated into the following geometry problem: let $S$ be a disc of radius $k_F$. Draw a vector $q \vec{e}_x$ starting from its center whose extremity defines the center of an other disc of the same radius called $S'$. Then, the maximal integration domain is $S'\setminus \{ S'\cap S\}$ which is $S'$ if $q>2k_F$ and has a crescent shape otherwise. We already see that $q=2k_F$ will play a major role in the process.

As a second step, the dirac distribution restricts the integration range to an even smaller area.
We treat the case $q>2k_F$ first. Eq.~(\ref{A1}) in this case reads
\begin{eqnarray}
\!\!\!\!\!\!\!\small{S_2(q,\omega)\!=\!\frac{1}{2\pi}\!\!\int_{S'}\!\!\!k\mathrm{d}k\!\!\!\int\!\!\mathrm{d}\theta_k\delta\!\left\{\!\omega\!-\!\frac{\hbar}{2m}[q^2\!+\!2kq\cos(\theta_k)]\!\!\right\}}
\end{eqnarray}
hence
\begin{eqnarray}
\frac{S_2(q,\omega)}{L_xL_y}=&&2\frac{1}{2\pi}\int_{(\omega-\omega_q)\frac{m}{\hbar q}}^{k_F}kdk\frac{1}{\frac{\hbar kq}{m}\sqrt{1-\left(\frac{\omega-\omega_q}{\hbar kq/m}\right)^2}}\nonumber\\
&&\Theta(\omega-\omega_{-,2})\Theta(\omega_{+,2}-\omega)
\end{eqnarray}
where the overall factor $2$ stems from the two allowed angles leading to the allowed value of the cosine. Then a simple integration yields the final result.
We now turn to the case $q<2k_F$. It is more convenient to use cartesian coordinates, and one has
\begin{eqnarray}
 \frac{S_2(q,\omega)}{L_xL_y}&&=\frac{1}{2\pi}\int_{S'\setminus \{ S'\cap S\}}\!\!\!\!\!\!\!\!\!\!dk_xdk_y\delta(\omega\!-\!\omega_q\!-\!k_x\hbar q/m)\nonumber\\
 &&=\frac{1}{2\pi}\int_{S'\setminus \{ S'\cap S\}}\!\!\!\!\!\!\!\!\!\!dk_xdk_y\frac{\delta(k_x\!-\!k_{x,0})}{\hbar q/m}\nonumber\\
 &&=\frac{1}{2\pi}\frac{m}{\hbar q}\int_{\{S'\setminus \{ S'\cap S\}\}\cap \{k_x=k_{x,0}\}}\!\!\!\!\!\!\!\!\!\!\!\!\!\!\!\!\!dk_y.
\end{eqnarray}
The last step consists in describing explicitly $\{S'\setminus \{ S'\cap S\}\}\cap \{k_x\!=\!k_{x,0}\}$, i.e. the intersection of a vertical straight line and a vertical crescent. If the line crosses the crescent in its filled part, i.e. if $k_F\!-\!q\leq k_{x,0}\leq k_F$, the domain is a straight line of length $2k_F|\sin(\theta_{k_F})|=2k_F\sqrt{1-\left(\frac{\omega-\omega_q}{qk_F\hbar/m}\right)^2}$. It corresponds to the case where $\omega_{+,2}\geq \omega\geq \omega_{-,2}$. If instead the line crosses the crescent and its hollow part, then the integration line is a segment deprived from a part in its inside, thus it consists in two lines. Then, the total length is $2k_F(|\sin(\theta_{k_F})|-|\sin(\theta_{k'})|)$, with
\begin{eqnarray}
 k_F|\sin(\theta_{k'})|&&=\sqrt{k_F^2-(q+k_F|\cos(\theta_{k_F})|)^2}\nonumber\\
 &&=k_F\sqrt{1-\left(\frac{q}{k_F}+\frac{\omega-\omega_q}{\hbar qk_F/m}\right)^2}\nonumber\\
 &&=k_F\sqrt{1-\left(\frac{\omega+\omega_q}{\hbar qk_F/m}\right)^2}.
\end{eqnarray}
The latter case implies that $\omega_{-,2}\geq \omega\geq 0$. Putting everything together and after some algebra, one finds the final result (Eq.~(\ref{Sd})):
\begin{eqnarray}
\label{A6}
S_{2}(q,\omega)&&=L_xL_y\frac{1}{\pi}\frac{m^2}{\hbar^2 q^2}\times \nonumber\\
&&\left\{\Theta(\omega_{-,2}-\omega)\Theta(2k_{F,2}-q)\right.\nonumber\\
&&\left[\sqrt{(\omega_{+,2}\!-\!\omega)(\omega\!+\!\omega_{-,2})}\!-\!\sqrt{(\omega_{-,2}\!-\!\omega)(\omega\!+\!\omega_{+,2})}\right]\nonumber\\
&&+\Theta(\omega_{+,2}-\omega)\Theta(\omega-\omega_{-,2})\nonumber\\
&&\left.\sqrt{[\omega_{+,2}-\omega][\omega-\rm{sign}(q-2k_{F,2})\omega_{-,2}]}\right\}.
\end{eqnarray}

We next calculate the drag force in the case $v<v_{F,2}$. Using Eq.~(\ref{explF}) we split the integral in three parts so that, up to a prefactor,
\begin{eqnarray}
&&F\!\!\propto\!\!\int_0^{q_{-}}\!\frac{dq}{q}\!\left[\!\sqrt{(\omega_{+}\!-\!qv)(qv\!+\!\omega_{-})}\!-\!\sqrt{(\omega_{-}\!-\!qv)(qv\!+\!\omega_{+}\!)}\right]\nonumber\\
&&\!+\!\!\int_{q_-}^{2k_F}\!\!\frac{dq}{q}\!\sqrt{\!(\omega_+\!-\!qv)(qv\!+\!\omega_{-})}\!+\!\!\!\int_{2k_F}^{q_+}\!\!\frac{dq}{q}\!\sqrt{\!(\omega_{-}\!-\!qv)(qv\!+\!\omega_{+}\!)}\nonumber\\
&&=\!\frac{\hbar}{2m}\!\!\left[\!\int_{0}^{q_+}\!\!\!\!dq\sqrt{\!(q\!+\!q_-)(q_+\!-\!q)}\!-\!\!\!\int_{0}^{q_-}\!\!\!\!\!\!dq\!\sqrt{\!(q\!+\!q_+)(q_-\!-\!q)}\!\right]
\end{eqnarray}
where $q_{\pm}\!\equiv\!\frac{2m}{\hbar}(v_F\!\pm\!v)$. The integrals can be expressed in terms of the hypergeometric function ${_2}F_1(1,-1/2;5/2;-x)$, using the property 3.196.1 of \cite{Gradshteyn}: $\int_0^u(x+\beta)^{\nu}(u-x)^{\mu-1}dx=\frac{\beta^\nu u^\mu}{\mu} {_2}F_1(1,-\nu;1+\mu;\frac{-u}{\beta})$, with $\beta\equiv q_{\pm}, u\equiv q_{\mp}, \nu\equiv 1/2, \mu\equiv 3/2$ here.
Then we use the following sequence of recursion theorems from the same reference to modify the arguments of the hypergeometric function: 9.137.1 with $\gamma=3/2, z=-x, \alpha=1$ and $\beta =-1/2$, 9.137.14 with $\beta=-1/2, \gamma=1/2, \alpha=1, z=-x$, 9.131.1, then 9.137.8 with $\alpha=0, \beta=1, \gamma=1/2$ and $z=\frac{x}{x+1}$, 9.131.1 again, twice, and eventually
\begin{equation}
 {_2}F_1(1/2,1/2;3/2;z^2)=\frac{\arcsin(z)}{z},
\end{equation}
which allows to express the result in terms of simple functions. Then, with the property
\begin{eqnarray}
 &&\arcsin(x)-\arcsin(y)\nonumber\\
 &&=\arcsin(x\sqrt{1-y^2}-y\sqrt{1-x^2}), xy>0
\end{eqnarray}
and putting back the prefactors, we find Eq.(\ref{F2}).

\subsection{Approach (II): direct calculation of $S_1$, dimensional crossover to $S_2$, direct calculation of $F_2$}
Recalling that the dynamical structure factor of a 1D Fermi gas reads
\begin{eqnarray}
 S_1(q,\omega)=\frac{\hbar}{m|q|}\Theta(\omega_{+}-\omega)\Theta(\omega-\omega_{-}),
\end{eqnarray}
then, for a 2D gas in a box of finite transverse size $L_y$ Eq.~(\ref{Equa5}) yields
\begin{eqnarray}
&&S_{Q1}(q\vec{e_x},\omega)=\sum_{k_y}\int dk_x\frac{L_x\hbar}{m|q|}\delta\left(\frac{\omega-\omega_{q}}{\hbar q/m}-k_{x}\right)\nonumber\\
&&\Theta(\epsilon_{F,y}-\epsilon_{k_x})\Theta(\epsilon_{k_x+q}-\epsilon_{F,y})\nonumber\\
&&=\frac{\hbar L_x}{m|q|}\sum_{k_y}\Theta(\omega_{+,y}-\omega)\Theta(\omega-\omega_{-,y})
\end{eqnarray}
where $\epsilon_{F,y}\equiv \epsilon_{F}\!-\!\frac{\hbar^2k_y^2}{2m}$, yielding Eq.~(\ref{induc}) in this peculiar case:
\begin{eqnarray}
 S_{Q1}(q\vec{e_x},\omega)=\frac{L_yk_F}{2\pi}\frac{1}{\tilde{M}}\sum_{j=-M}^MS_1(q\vec{e_x},\omega;\sqrt{1-(j/M)^2}).
\end{eqnarray}
In reduced units where $q/k_F\to q$ and $\omega/\omega_F\to \omega$, in the limit $M\to +\infty$ where the Riemann sum becomes an integral we are left to compute
\begin{eqnarray}
\!\!\!\!\!\!\!\!\!\!\!\!\small{I\!\!\equiv\!\!\int_0^1\!\!\!\!dx\Theta(\!q^2\!\!+\!2q\sqrt{1\!-\!x^2}\!-\!\omega\!)\Theta(\!\omega\!-\!|q^2\!\!-\!2q\sqrt{1\!-\!x^2}|\!)}\!.
\end{eqnarray}
Special attention should be paid to the integration range, namely the argument of the Heaviside distributions and the square roots must be non-negative. In the discussion, it is useful to consider separately the cases $q>2$ and $q<2$, but also $\omega>q^2$ and $\omega<q^2$, which appear naturally. After a careful analysis one finds :
\begin{eqnarray}
&&I\!\!=\!\!\Theta(q^2\!-\!\omega)\Theta(q\!-\!2)[\Theta(\omega_+\!-\!\omega)]\Theta(\omega\!-\!\omega_-)\!\!\!\int_0^{\sqrt{\!1\!-\!\left(\!\frac{q^2\!-\!\omega}{2q}\!\right)^2}}\!\!\!\!\!\!dx\nonumber\\
&&+\Theta(q^2\!-\!\omega)\Theta(2\!-\!q)[\Theta(\omega_+\!-\!\omega)]\!\!\int_{\sqrt{1-\left(\frac{q}{2}\right)^2}}^{\sqrt{\!1\!-\!\left(\frac{q^2\!-\!\omega}{2q}\right)^2}}dx\nonumber\\
&&+\Theta(q^2\!-\!\omega)\Theta(2\!-\!q)[\Theta(\omega_+\!-\!\omega)]\Theta(\omega\!-\!\omega_-)\!\!\int_0^{\sqrt{\!1\!-\!\left(\frac{q}{2}\right)^2}}dx\nonumber\\
&&+\Theta(q^2\!-\!\omega)\Theta(2\!-\!q)[\Theta(\omega_+\!-\!\omega)]\Theta(\omega_-\!-\!\omega)\!\!\int_{\sqrt{\!1\!-\!\left(\!\frac{\omega\!+\!q^2}{2q}\!\right)^2}}^{\sqrt{1\!-\!\left(\!\frac{q}{2}\!\right)^2}}\!\!dx\nonumber\\
&&+\Theta(\omega\!-\!q^2)\Theta(q\!-\!2)\Theta(\omega_+\!-\!\omega)\Theta(\omega\!-\!\omega_-)\!\!\!\int_{0}^{\sqrt{\!1\!-\!\left(\!\frac{\omega\!-\!q^2}{2q}\!\right)^2}}\!\!dx\nonumber\\
&&+\Theta(\omega\!-\!q^2)\Theta(2\!-\!q)\Theta(\omega_+\!-\!\omega)\!\!\int_{\sqrt{1-\left(\frac{q}{2}\right)^2}}^{\sqrt{\!1\!-\!\left(\frac{\omega\!-\!q^2}{2q}\right)^2}}dx \nonumber\\
&&+\Theta(\omega\!-\!q^2)\Theta(2\!-\!q)\Theta(\omega_+\!-\!\omega)\Theta(\omega\!-\!\omega_-)\!\!\int_{0}^{\sqrt{\!1\!-\!\left(\frac{q}{2}\right)^2}}dx \nonumber\\
&&+\Theta(\omega\!-\!q^2)\Theta(2\!-\!q)\Theta(\omega_+\!\!-\!\omega)\Theta(\omega_-\!\!-\!\omega)\!\!\!\int_{\sqrt{\!1\!-\!\left(\!\frac{\omega\!+\!q^2}{2q}\!\!\right)^2}\!}^{\sqrt{\!1\!-\!\left(\frac{q}{2}\right)^2}}dx.\nonumber\\
\end{eqnarray}
These terms can be recombined pairwise, thus recovering Eq.~(\ref{A6}). The drag force is then computed directly as in (I).

\subsection{Approach (III): direct calculation of $S_1$ and $F_1$, dimensional crossover to $F_2$}
Given $S_1$, $F_1$ is readily computed, treating separately the cases $v\!>\!v_{F,1}$ and $v\!<\!v_{F,1}$. For a q1D system, we get
\begin{eqnarray}
 &&F_{Q1}(v)=\frac{U^2}{2\pi\hbar L_xL_y}\int_0^{+\infty}\!\!dq S_{q1D}(q\vec{e_x},qv)q\nonumber\\
 &&=\!\frac{U^2}{2\pi\hbar}\frac{m}{\hbar}\!\!\int_0^{+\infty}\!\!\!\!\!\!dq\!\sum_{k_y}\Theta(\omega_{+,y}\!-\!qv)\Theta(qv\!-\!\omega_{-,y})\!.
\end{eqnarray}
Interchanging the sum and the integral, we find
\begin{eqnarray}
&&F_{Q1}(v)=\frac{1}{L_y}\frac{k_{F}}{2\pi}\frac{2U_{2}^2m}{\hbar^2}\nonumber\\
&&\sum_{j=-M}^M\!\!\left\{\frac{v}{v_{F}}\Theta[v_{F}(j)\!-\!v]\!+\!\frac{v_{F}(j)}{v_{F}}\Theta[v\!-\!v_{F}(j)]\right\}
\end{eqnarray}
where $v_{F}(j)\equiv v_F\sqrt{1-j^2/\tilde{M}^2}$, so that in the limit $M\to +\infty$, using reduced variables we have to compute
\begin{eqnarray}
\!\!\!\!\!\!\!\!J\!\equiv\!\!\!\int_0^1\!\!\!dx[u\Theta(\sqrt{\!1\!-\!x^2}\!-\!u)\!\!+\!\!\sqrt{\!1\!-\!x^2}\Theta(u\!-\!\sqrt{\!1\!-\!x^2})]
\end{eqnarray}
which readily yields Eq.~(\ref{F2}) in the main text.

\section{Drag force in $d$ dimensions}
\label{Fd}
In this Appendix we provide the technical details on the derivation of the drag force in arbitrary integer dimension $d$, Eq.~(\ref{Fdd}). We start from Eq.~(\ref{Sd}) combined with Eq.~(\ref{explF}). If $v>v_{F,d}$, one easily shows that, up to a prefactor,
\begin{eqnarray}
 F_d(v)\propto \int_{q_{-,d}}^{q_{+,d}}\mathrm{d}q [(q_{+,d}-q)(q-q_{-,d})]^{\frac{d-1}{2}}
\end{eqnarray}
where $q_{\pm,d}\equiv \frac{2m}{\hbar}(v\pm v_{F,d})$. Using the property
\begin{eqnarray}
\!\!\!\int_a^b \!\!\mathrm{d}x (x-a)^{\mu-1}(b-x)^{\nu-1}\!=\!(b-a)^{\mu+\nu-1}\!B(\mu,\nu)
\end{eqnarray}
where $B$ is the Euler Beta function and $B(x,x)=2^{1-2x}\Gamma(1/2)\Gamma(x)/\Gamma(x+1/2)$, the problem is readily solved by expliciting the prefactor, yielding
\begin{eqnarray}
 F_d(v>v_{F,d})=\frac{2U_d^2mn_d}{\hbar^2}.
\end{eqnarray}
The case $v<v_{F,d}$ is more involved. In this case $q_{\pm,d}\equiv \frac{2m}{\hbar}(v_{F,d}\pm v)$, and one can show that, up to a prefactor,
\begin{eqnarray}
\label{arg}
F_d(v<v_{F,d})&&\propto\int_0^{q_{+,d}}\!\!\mathrm{d}q[(q_{+,d}-q)(q+q_{-,d})]^{\frac{d-1}{2}}\nonumber\\
&&-\int_0^{q_{-,d}}\!\!\mathrm{d}q[(q_{-,d}-q)(q+q_{+,d})]^{\frac{d-1}{2}}.
\end{eqnarray}
Using the identity \cite{Gradshteyn}
\begin{eqnarray}
\!\!\!\!\!\!\!\int_0^u\!\!\!\!\!dx(x\!+\!\beta)^{\nu}\!(u\!-\!x)^{\mu\!-\!1}\!dx\!=\!\frac{\beta^{\nu}\!u^{\mu}}{\mu}\!{_2}F_1\!\!\left(\!1,\!-\!\nu;1\!+\!\mu;\!-\frac{u}{\beta}\!\right)\!\!,
\end{eqnarray}
where ${_2}F_1$ is the hypergeometric function, we obtain Eq.~(\ref{Fdd}). Although the hypergeometric representation is convenient to synthetize the results, we would like to express the drag force in terms of simple functions. 

Using properties of hypergeometric functions, in odd dimensions one finds
\begin{eqnarray}
&&\frac{F_d(v<v_F)}{F_d(v_F)}=1+\frac{2}{\sqrt{\pi}}\frac{\Gamma(\frac{d+2}{2})}{\Gamma(\frac{d+1}{2})}(1-u_d^2)^{\frac{d-1}{2}}(1+u_d)\frac{1}{d}\times\nonumber\\
&&\!\left(\!1\!\!+\!\!\sum_{j=1}^{\frac{d-1}{2}}\!\left(1\!-\!u_d\right)^{-\!j}\!\prod_{i=1}^j\!a_{d,i}\!-\!\left(1\!-\!\!u_d\!\right)^{\frac{1\!-\!d}{2}}\!\!\!\left(\!\frac{2}{1\!+\!u_d\!}\!\right)^{\!\!\!\frac{d\!+\!1}{2}}\!\prod_{i=1}^{\frac{d\!-\!1}{2}}\!a_{d,i}\!\!\right)
\end{eqnarray}
where $a_{d,i}\equiv \frac{d\!+\!1\!-\!2i}{d\!-\!i}$.

In the even dimension case, the expression is a bit more involved. To obtain it we showed that for any integer $J$,
\begin{eqnarray}
 &&{_2}F_1\!\!\left(\!1,k\!+\!1;\frac{3}{2};x\!\right)\nonumber\\
 &&\!=\!{_2}F_1\!\!\left(\!1,k\!-\!J;\frac{3}{2};x\!\right)\!a_J\!+\!{_2}F_1\!\!\left(\!1,k\!-\!J\!-\!1;\frac{3}{2};x\!\right)\!b_J
\end{eqnarray}

where $a_{-1}\!=\!1$, $b_{-1}\!=\!0$, $a_j=c_ja_{j-1}+b_{j-1}$, $b_j=d_ja_{j-1}$, $c_j\equiv \frac{4(k-j)-3+2(1-(k-j))x}{2(k-j)(1-x)}$, $d_j\equiv \frac{3-2(k-j)}{2(k-j)(1-x)}$.

The cut-off index $J$ is chosen so that $J=k-1$, yielding after some algebra

\begin{eqnarray}
&& \frac{F_d(v<v_F)}{F_d(v_F)}=1+\frac{2}{\sqrt{\pi}}\frac{\Gamma(\frac{d+2}{2})}{\Gamma(\frac{d+1}{2})}(1-u_d^2)^{\frac{d-1}{2}}(1+u_d)\frac{1}{d}\times\nonumber\\
&&\!\!\left(\!\!1\!\!+\!\!\!\!\sum_{j=1}^{\frac{d\!-\!2}{2}}(1\!-\!u_d)^{\!-\!j}\!\prod_{i\!=\!1}^ja_{d,i}\!-\!(1\!-\!u_d)^{\frac{2\!-\!d}{2}}\!\prod_{i=1}^{\frac{d\!-\!2}{2}}\!a_{d,i}\!\!\left[2a_{\frac{d\!-\!2}{2}}f(u_d)\!+\!b_{\frac{d\!-\!2}{2}}\!\right]\!\!\right)\nonumber\\
&&f(u)\equiv \frac{\arcsin\!\left(\!\sqrt{\frac{1-u}{2}}\right)}{\sqrt{1-u^2}},
\end{eqnarray}
where a product with upper index $0$ is equal to $1$ by convention.

The same techniques allowed us to find Eqs.~(\ref{Fsup}) and (\ref{Finf}). Using the property
\begin{eqnarray}
 &&\int_0^1dx x^{\lambda-1}(1-x)^{\mu-1}(1-\beta x)^{-\nu}\nonumber\\
 &&=B(\lambda,\mu){_2}F_1(\nu,\lambda;\lambda+\mu;\beta)\nonumber\\
 &&Re(\lambda)>0, Re(\mu)>0, |\beta|<1,
\end{eqnarray}
six of the Gauss recursion theorems and
\begin{eqnarray}
 {_2}F_1(1/2,1;3/2;-x^2)=\frac{\arctan(x)}{x^2}
\end{eqnarray}
we evaluated
\begin{eqnarray}
&&\int_0^1 \mathrm{d}x\sqrt{x}\sqrt{1-\beta x}, |\beta|<1\nonumber\\
&&=\frac{1}{4\beta}\left[\frac{\arcsin(\sqrt{\beta})}{\sqrt{\beta}}+\!(2\beta\!-\!1)\sqrt{1\!-\!\beta}\right]\!.
\end{eqnarray}

%%%%%%%%%%%%%%%%%%%%%%%%%%%%%%%%%%%%%%%%%%%%%%%%%%%%%%%%%%%%%%%%%%%%%%%%%%%%%%%%%%%%%%%%%%%%%%%%%%%%%%%%%%%%%%%%%%%%%%%


\begin{thebibliography}{99}
%%%%%%%%%%%%%%%%%%%%%%%%%%%%%%%%%%%%%%%%%%%%%%%%%%%%%%%%%%%%%%%%%%%%%%%%%%%%%%%%%%%%%%%%%%%%%%%%%%%%%%%%%%%%%%%%%%%%%%%

\bibitem{Bloch}
M. Greiner, O. Mandel, T. Esslinger, T. W. H\"ansch, and I. Bloch, Nature \textbf{415}, 39 (2002)
\bibitem{Fleischer}
J.W. Fleischer, M. Segev, M.K. Efremidis, D.N. Christodoulides, Nature \textbf{422},147-150 (2003)
\bibitem{Hadzi}
Z. Hadzibabic, P. Kr\"uger, M. Cheneau, B. Battelier, and J. Dalibard, Nature \textbf{441}, 1118-1121 (2006)
\bibitem{Paredes}
B. Paredes, A. Widera, V. Murg, O. Mandel, S. F\"olling, I. Cirac, G.V. Shlyapnikov, T.W. H\"ansch, and I. Bloch, Nature \textbf{429}, 277 (2004)
\bibitem{VanDruten}
W. Ketterle and N.J. van Druten, Phys. Rev. A \textbf{54}, 656 (1996)
\bibitem{Roati}
G. Roati et al., Nature \textbf{453}, 895 (2008)
\bibitem{Billy}
J. Billy et al., Nature \textbf{453}, 891 (2008)
\bibitem{Esslinger}
T. St\"oferle, H. Moritz, C. Schori, M. K\"ohl, and T. Esslinger, Phys. Rev. Lett. \textbf{92}, 13 (2004)
\bibitem{Chen}
Y.-A. Chen, S. D. Huber, S. Trotzky, I. Bloch, and E. Altman, Nature Physics \textbf{7}, 61 (2011)
\bibitem{Naegerl}
E. Haller, M. Gustavsson, M.J. Mark, J.G. Danzl, R. Hart, G. Pupillo, H.-C. N\"agerl, Science \textbf{325}, 5945 (2009)
\bibitem{Dalibard}
R. Desbuquois, L. Chomaz, T. Yefsah, J. L\'eonard, J. Beugnon, C. Weitenberg and J. Dalibard, Nature Physics \textbf{8}, 645-648 (2012)
\bibitem{Widera}
A. Widera, S. Trotzky, P. Cheinet, S. F\"olling, F. Gerbier, I. Bloch, V. Gritsev, M. D. Lukin, and E. Demler, Phys. Rev. Lett. \textbf{100}, 140401 (2008)
\bibitem{Gorlitz}
A. G\"orlitz, J.M. Vogels, A.E. Leanhardt, C. Raman, T.L. Gustavson, J.R. Abo-Shaeer,
A.P. Chikkatur, S.Gupta, S.Inouye, T.Rosenband, and W. Ketterle, Phys. Rev. Lett. \textbf{87}, 13 (2001)
\bibitem{Chung}
M.-C. Chung and A. B. Bhattacherjee, Phys. Rev. Lett. \textbf{101}, 070402 (2008)
\bibitem{Sandler}
L.P. Sandler and D.L. Maslov, Phys. Rev. B \textbf{55}, 13808 (1997)
\bibitem{Crepin}
F. Crepin, N. Laflorencie, G. Roux and P. Simon, Phys. Rev. B \textbf{84}, 054517 (2011)
\bibitem{Gogolin}
R. Egger and A.O. Gogolin, Phys. Rev. Lett. \textbf{79}, 25 (1997)
\bibitem{Killi}
M. Killi, T.-Z. Wei, I. Affleck, and A. Paramekanti, Phys. Rev. Lett. \textbf{104}, 216406 (2010)
\bibitem{Sommer}
A.T. Sommer, L.W. Cheuk, M.J.H. Ku, W.S. Bakr, and M.W. Zwierlein, Phys. Rev. Lett. \textbf{108}, 045302 (2012)
\bibitem{Joung}
D. Joung and S.I. Khondaker, Phys. Rev. B \textbf{89}, 245411 (2014)
\bibitem{Vogler}
A. Vogler, R. Labouvie, G. Barontini, S. Eggert, V. Guarrera, and H. Ott, Phys. Rev. Lett. \textbf{113}, 215301 (2014)
\bibitem{Stenger}
J. Stenger, S. Inouye, A. P. Chikkatur, D. M. Stamper-Kurn, D. E. Pritchard, and W. Ketterle, Phys. Rev.
Lett. \textbf{82}, 4569-4573 (1999)
\bibitem{Ozeri}
R. Ozeri, N. Katz, J. Steinhauer and N. Davidson, Rev. Mod. Phys. \textbf{77}, 187 (2005)
\bibitem{Calabrese}
J.-S. Caux and P. Calabrese, Phys. Rev. A \textbf{74}, 031605 (2006)
\bibitem{Brunello}
A. Brunello, F. Dalfovo, L. Pitaevskii, S. Stringari, and F. Zambelli, Phys. Rev. A \textbf{64}, 063614 (2001)
\bibitem{David}
N. Fabbri, M. Panfil, D. Cl\'ement, L. Fallani, M. Inguscio, C. Fort and J.-S. Caux, Phys. Rev. A \textbf{91}, 043617 (2015)
\bibitem{Meinert}
Florian Meinert, Milosz Panfil, Manfred J. Mark, Katharina Lauber, Jean-S\'ebastien Caux, Hanns-Christoph N\"agerl, Phys. Rev. Lett.  \textbf{115}, 085301 (2015)
\bibitem{Brand}
J. Brand and A.Y. Cherny, Phys. Rev. A \textbf{72}, 033619 (2005)
\bibitem{Pitaevskii}
G.E. Astrakharchik and L.P. Pitaevskii, Phys. Rev. A \textbf{70}, 013608 (2004)
\bibitem{Ketterle}
R. Onofrio, C. Raman, J. M. Vogels, J. R. Abo-Shaeer, A. P. Chikkatur, and W. Ketterle, Phys. Rev. Lett. \textbf{85}, 11 (2000)
\bibitem{Singh}
V. P. Singh, W. Weimer, K. Morgener, J. Siegl, K. Hueck, N. Luick, H. Moritz, and L. Mathey, arXiv:1509.02168v1 [cond-mat.quant-gas] (2015)
\bibitem{Pines}
P. Nozi\`eres and D. Pines, The Theory of Quantum Liquids: Superfluid Bose Liquids (Addison-Wesley, Redwood City, 1990)
\bibitem{Fetter}
A.L.Fetter and J.D. Waleczka, Quantum Theory of Many-Particle Systems (McGraw-Hill, 2003)
\bibitem{Vignolo}
P. Vignolo, A. Minguzzi and M.P. Tosi, Phys. Rev. A \textbf{64}, 023421 (2001)
\bibitem{Tomonaga}
S. Tomonaga, Prog. Theor. Phys \textbf{5}, 544 (1950)
\bibitem{Mattis}
D.C. Mattis and E.H. Lieb, J. Math. Phys. \textbf{6}, 304 (1965)
\bibitem{Haldane}
F.D.M. Haldane, J. Phys. C: Solid State Phys. \textbf{14}, 2585 (1981)
\bibitem{Cazalilla}
M.A. Cazalilla, J. Phys. B: At. Mol. Opt. Phys. \textbf{37}, S1-S47 (2004)
\bibitem{Giamarchi}
T. Giamarchi, Quantum Physics in One Dimension (Oxford University Press, 2004)
\bibitem{Voit}
J. Voit, Rep. Prog. Phys. \textbf{58}, 977 (1995)
\bibitem{Hadzibabic}
A.L. Gaunt, T.F. Schmidutz, I. Gotlibovych, R.P. Smith, and Z. Hadzibabic, Phys. Rev. Lett. \textbf{110}, 200406 (2013)
\bibitem{Ashcroft}
N. W. Ashcroft and N. D. Mermin, Solid State Physics (Brooks/Cole, Pacific Grove, CA, 1976)
\bibitem{Cherny}
A.Y. Cherny, J.-S. Caux and J. Brand, Front. Phys. \textbf{7}(1): 54-71 (2012)
\bibitem{Golovach}
V.N. Golovach, A. Minguzzi and L.I. Glazman, Phys. Rev. A \textbf{80}, 043611 (2009)
\bibitem{Lang}
G. Lang, F. Hekking and A. Minguzzi, Phys. Rev. A \textbf{91}, 063619 (2015)
\bibitem{Metzner}
C. Castellani, C. Di Castro, and W. Metzner, Phys. Rev. Lett. \textbf{72}, 3 (1994)
\bibitem{Wen}
P.-A. Bares and X.-G. Wen, Phys. Rev. B \textbf{48}, 12 (1993)
\bibitem{Bartosch}
L. Bartosch and P. Kopietz, Phys. Rev. B \textbf{59}, 8 (1998)
\bibitem{Iucci}
A. Iucci, G.A. Fiete and T. Giamarchi, Phys. Rev. B \textbf{75}, 205116 (2007)
\bibitem{Orignac}
E. Orignac and R. Citro, J. Stat. Mech., 12020 (2012)
\bibitem{Guinea}
F. Guinea, G.Zimanyi, Phys. Rev. B \textbf{47}, 1 (1993)
\bibitem{Imambekov}
A. Imambekov, T.L. Schmidt, and L.I. Glazman, Rev. Mod. Phys. \textbf{84}, 1253 (2012)
\bibitem{Gradshteyn}
I.S. Gradshteyn, I.M. Ryzhik, Table of Integrals, Series and Products (Academic Press, $6^{th}$ edition, 2000)


\end{thebibliography}
\end{document}